Comment on Quantized acoustoelectric current transport through a static quantum dot using a surface acoustic wave.

V. I. Talyanskii
*Cavendish Laboratory, University of Cambridge, Madingley Road, Cambridge CB3 0HE, United Kingdom*

Recently a new explanation for the quantized current of the surface acoustic wave (SAW)-driven single electron pumps was suggested [1]. In the SAW-driven pumps the electrons are transported by the SAW along a one-dimensional semiconductor channel. The suggested in ref.1 mechanism requires the presence of a static quantum dot in the 1D channel and is qualitatively different from that described in the original papers on single electron SAW-pumps.
We have analysed the available experimental data on the quantized current (including the data presented in ref.1) and have come to the conclusion that the data can not be explained by the mechanism suggested in ref.1. Moreover, we argue that the data presented in ref.1 don't warrant the claim of the authors that a new mechanism of the current quantization was observed.

## I. Introduction

In a recent paper [1], the authors claimed the observation of a new mechanism of quantization of the current induced by a surface acoustic wave (SAW) in a one-dimensional channel that contained an impurity-induced quantum dot (QD). Whilst the mechanism was claimed to be observed in the domain of low SAW power levels, the authors of ref.1 suggested that the QD is relevant to the previous observations of the quantized acoustoelectric current [2-6]. We shall refer to the suggested in ref. 1 mechanism as to the SAW-turnstile (ST) mechanism, because it is similar to the single electron turnstile that was studied in refs. [7, 8].
The claim of ref.1 immediately raises serious questions.
Firstly, the turnstile single electron devices [7, 8] feature very specific dependences of the current on source drain bias and on the voltage on the gate that controls the position of the QD spectrum with respect to the Fermi levels of the leads. Those dependences are fundamentally different from and in apparent disagreement with the experimental dependences of the current produced by the SAW single electron pumps [2-6].
Secondly, the situation when the quantization of the acoustoelectric current is observed, and at the same time no impurity-induced QD in the channel is detected is rather common. One example such situation shall be illustrated below (Figs. 5 and 7). The opposite situation when the impurity defined dot is present, however no quantization of the acoustoelectric current is observed is also common and is illustrated in Fig. 2 below.
The ST mechanism [1] implies that the SAW charge transport in quasi one-dimensional semiconductor channels is extremely strongly affected by impurities present in or nearby of the channel. So the ST mechanism, if true, has important consequences for the development of the SAW-based current standard and for the use of the SAW charge transport in quantum information processing applications [9, 10]. In particular, the SAW quantum computing proposal [10] will be rendered not viable by the frequent captures of the SAW driven electrons by the unavoidable impurity centres.
In this Comment we investigate (*i*) whether the ST model is relevant to the previous research [2-6], and (*ii*) whether the experimental data presented in ref.1 warrant the claims made in that publication. We conclude that all the experimental evidence concerning the quantized acoustoelectric current (including that in ref.1) is in contradiction with the ST mechanism.
Authors of ref.1 have not discussed in detail the crucial corollaries of the ST model that would allow assessing the relevance of the model. Also the relevant aspects of the previous studies have not been referred to. We shall compensate for those omissions by outlining the important implications of the ST model and by presenting the relevant details of the previous studies [2-6].
The Comment is organised as follows. We shall present the relevant details of the previous studies, and then outline the essential corollaries of the ST model and compare them with the previous experiments [2-6]. Finally the experimental evidence presented in ref.1 will be examined.

## II. The quantization of the acoustoelectric current

In the SAW pumps the SAW drives electrons along a quasi one-dimensional channel (shown by the dashed lines in Fig.1a) formed in a GaAs heterojunction by a split gate [2, 3]. The acoustoelectric current (graph 1 in Fig.1b) reveals the plateaux at the current values *ef*, 2*ef*, and so on, where *e* is the electron charge and *f* is the SAW frequency.

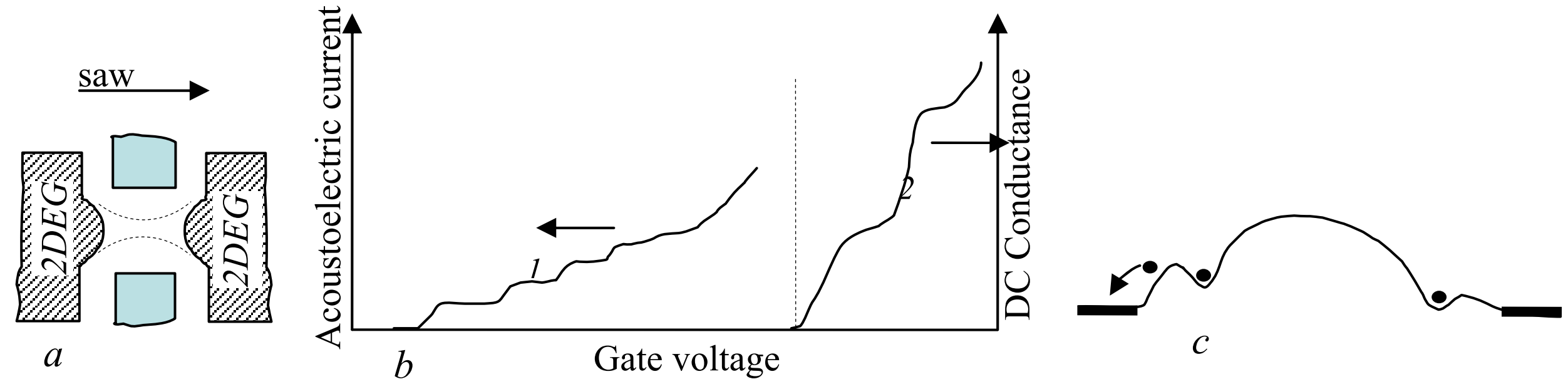

Fig. 1.
*a*. Schematic of the 1D channel along which the SAW drives electrons. *b*. Acoustoelectric current and the channel conductance vs gate voltage. The panel illustrates that the quantized current is observed in the depleted channel. *c*. The potential profile in the channel in the presence of the SAW. The panel illustrates the mechanism of current quantization that was suggested in refs [2, 3].

The explanation of the current quantization (we shall refer to it as to the original model) was based on the picture of the potential minimum (Fig.1c) moving up the potential hill that formed by the voltage applied to the split gate [2, 3]. The shape of the minimum is determined by the superposition of the travelling SAW potential and the static potential due to the gate, and it changes as the minimum moves up the potential hill. A number of electrons that the minimum can transfer from the source to the drain 2DEG is controlled by the steepest part of the static potential and by the amplitude of the SAW potential. In the experiments [2-6] the quantized current has been observed in the domain of the gate voltages below the conductance (measured with the SAW off; curve 2 in Fig.1b) pinch off. The fact that the current quantization has been observed only in the closed (pinched off) channel is relevant in the context of this Comment.

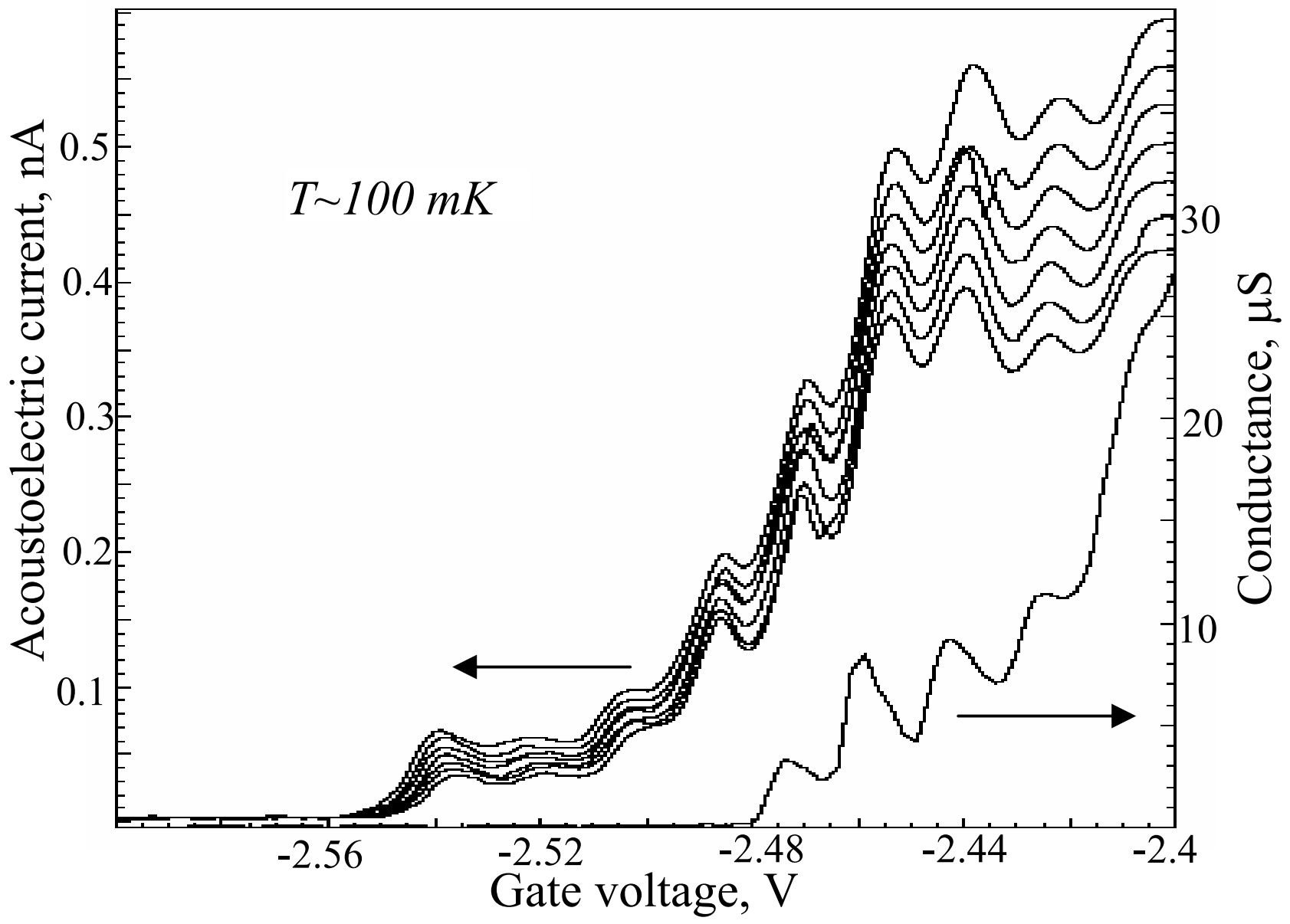

Fig.2 The acoustoelectric current for different SAW powers (see text) and the channel's conductance (in the absence of the SAW) vs gate voltage.

The authors of ref.1 state (we quote): "*The device used in this work was geometrically identical to those used in previous studies with the exception that the channel region of this device contained a quantum dot formed by an impurity potential near the two-dimensional electron gas*".

The second part of that statement is not strictly correct. The observation of the impurity-induced QD and its relevance to the current quantization had been discussed in refs. [2, 4-6].

An impurity-induced dot manifests itself by the characteristic Coulomb blockade peaks in the dc conductance and the acoustoelectric current vs gate voltage graphs [2, 4-6]. It had been observed that the impurity peaks in the conductance are accompanied by the peaks (not by the plateaus) in the acoustoelectric current. That observation led us to suggest [2, 4, 6] that the impurity potential is not essential for appearance of the quantized current, and that the impurities within or nearby of the channel have a deteriorating effect on the accuracy of the current quantization. Presence of the impurity-induced QD in the channel may lead to the more complicated profile of the static potential as compared with that shown in Fig. 1c. A profile with two maxima (corresponding to the two potential barriers defining the QD) is then expected. According to the original model [1, 2] the quantization of the acoustoelectric current is still possible if the SAW field is strong enough to provide a moving potential minimum even at the steepest parts of the static potential.

In practice, the devices revealing both the current quantization and the Coulomb peaks, or either the quantization or the peaks had been observed [2-6]. Fig. 2 illustrates the dependences of the acoustoelectric current and the dc conductance vs gate voltage for the channel containing an impurity-induced QD. The current graphs correspond to different rf powers from +6.25dBm (for the bottom graph) up to +8dBm incrementing in 0.25 dBm steps (similar dependences were observed in this sample in the range from -8dBm to +8dBm). It is seen that the Coulomb blockade peaks in the conductance are reflected by peaks (rather than the quantization plateaus) in the acoustoelectric current.

III. Dependence of the acoustoelectric current on the SAW amplitude, dc bias, and the gate voltage in the ST model.

For convenience we first outline the basic features of the ST model, as they presented in ref.1. A QD which is formed by a superposition of the potential induced by the split gate and that of a nearby impurity is subjected to the action of the SAW (Fig.3). It was suggested [1] that the electrochemical potentials of the adjacent 2DEGs are unchanged by the SAW, and that the SAW modulates the

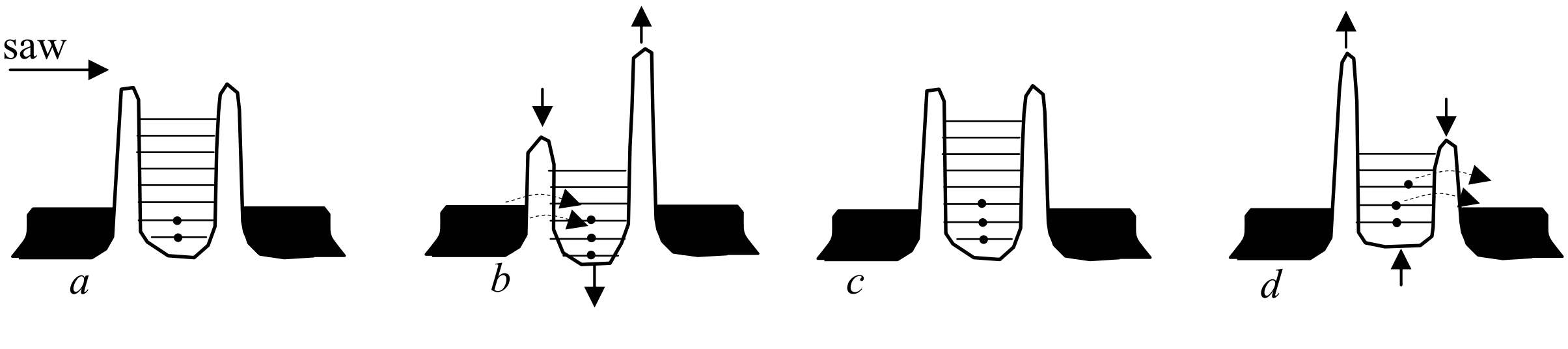

Fig. 3 Panels *a* to *d* illustrate the electron transport through the static quantum dot in the ST model. See text for detail.

barriers heights and the position of the bottom of the conduction band. Throughout the discussion by the energy levels in the dot we mean the levels that can be occupied by the interacting electrons. Those level are approximately separated by the Coulomb charging energy $\varepsilon_c$. On the energy scale, the barriers height and the bottom of the conduction band oscillate with amplitude $eA_{saw}$, where $A_{saw}$ is the amplitude of the SAW electrostatic potential. It was also assumed [1] that the phase shifts between the oscillations of the barriers and the states in the dot are such that when the bottom of the conduction band in the dot (Fig.3b) moves down the entry barrier (the one on the side from which the SAW comes) is opening and the exit barrier is closing. When the bottom of the conduction band moves up, the exit barrier gets open and the entry one is getting closed (Fig.3d). We shall analyse the corollaries of the ST model by using an approximation which captures the main assumptions of the ST model and features the quantized electron transport. In that approximation (which we shall keep refer to as the ST model) half of SAW period the entry barrier is open, the exit barrier is closed, and the bottom of the conduction band in the dot is lower than its equilibrium position (when the SAW is off) by $eA_{saw}$ (Fig.3b). During the second half of the SAW period the bottom of the conduction band is above its equilibrium position by $eA_{saw}$, the entry barrier is closed, and the exit barrier is open (Fig.3d). We note that the use of the above approximation does not change the conclusions, which are in fact "predetermined" by the assumption of the turnstile mechanism of the current quantization.
All the predictions of the ST model could be readily extracted from the Kowenhoven *et. al.* work [7, 8] on the turnstile single electron device, by noticing that the SAW amplitude in the ST model plays the role (in addition to the modulation the barriers' height) of the dc bias in the turnstile mechanism in refs [7, 8].
Fig. 4a shows the expected (within ST model) dependence of the current on the SAW amplitude (the red curve) in the absence of source-drain bias $U_{sd}=0$. The shape of the graph depends on the position of the Fermi level with respect to the dot's spectrum (controlled by the gate voltage $V_g$). When $\varepsilon_F$ is in the middle between two dot's levels then the current plateaux corresponding to 2, 4, 6 and so on electrons transferred per cycle are expected. If $\varepsilon_F$ is aligned with energy level in the dot then the plateaux corresponding to 1, 3, 5 and so on transferred electrons should be seen. This is because two additional levels in the dot are involved into play simultaneously as the SAW amplitude increases [1]. If $\varepsilon_F$ is not in a symmetrical position in respect to the dot's spectrum, then all the current plateaux should be visible, with the widths of the plateaux changing in some periodic fashion.
This is illustrated in Fig.4a for the situation when the 2DEG's Fermi level $\varepsilon_F$ is slightly above the highest filled (in the absence of the SAW) energy level in the dot. In Fig.4a the horizontal dashed line represent the energy levels in the dot separated by the interval $\varepsilon_c$, and the two tilted (at 45 degrees) lines serve to define the values of $A_{saw}$ at which an additional level in the dot starts to take part in the electron transport. It is seen that the odd current plateaux (1, 3, 5 and so on) are wider than the even ones. When $\varepsilon_F$ is closer to the middle between the dot's levels the even plateaux should dominate.
The dependence of the acoustoelectric current on the source-drain bias has the universal step-like form [7, 8] with the current increasing by *ef* with the increase of the bias by $\varepsilon_C/e$ (Fig.4b) The acoustoelectric current can be estimated (with the accuracy of ± *ef*) as $e^2f(2A_{saw}+U_{sd})/\varepsilon_c$. In Fig.4 positive value of the current corresponds to the electron flow in the direction of the SAW

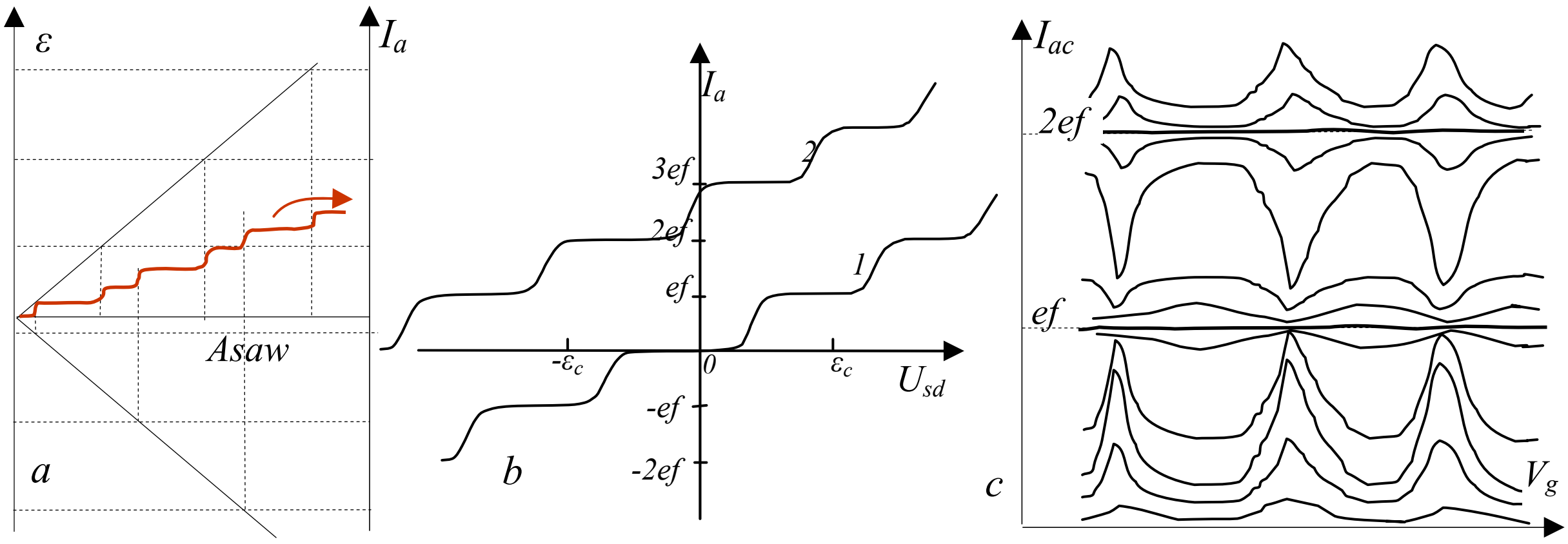

Fig.4 The acoustoelectric current as a function of the SAW amplitude (panel *a*), source-drain voltage (*b*), and gate voltage (*c*). The dependences are the direct consequence of ST model. See text for details.

propagation,and positive values of $U_{sd}$ corresponds a higher Fermi level of the source 2DEG with respect to the drain 2DEG. Position of the graph in the $I_a$, $U_{sd}$-plane depends on $A_{saw}$ and $V_g$. In Fig.4b graph 1 corresponds to $A_{saw}e<\varepsilon_c/2$ and the 2DEGs' Fermi level (at $U_{sd}$=0) in the middle between two dot's levels; graph 2 corresponds to $A_{saw}e\geq\varepsilon_c$ and 2DEGs' Fermi level (at $U_{sd}$=0) aligned with an energy level in the dot. It is important that the two parameters determining the graphs' shape, the distance between the steps ($\varepsilon_c$), and the steps height (*ef*), are usually well known ($\varepsilon_c$ is measured independently from the dc *I-V* characteristic of the dot).
Fig.4c illustrates the $I_a(V_g)$ dependences for different values of $A_{saw}$ and $U_{sd}=0$. In the ST model the voltage applied to the gate controls the position of the energy levels in the dot relatively to the Fermi level of the 2DEG. This immediately implies a periodic dependence of the relevant quantities on the gate voltage (provided a quasi equidistant distribution of the levels in the dot that does not strongly depend on the gate voltage; the condition which is usually satisfied). As gate voltage changes, the number of the dot's states in the interval $2eA_{saw}$ oscillates resulting in the oscillating behaviour of the current. For example, if $2eA_{saw}<\varepsilon_c$, in some interval of the gate voltage there will be no dot's state in the energy interval $\varepsilon_F\pm eA_{saw}$ and no or a small (compared with $ef$) current flows across the dot. So for $2eA_{saw}<\varepsilon_c$, the current will oscillate between 0 and $ef$. When, $2eA_{saw}=\varepsilon_c$ there is always one and only one dot's level in the interval $2eA_{saw}$, and the current stays constant and equal to the quantized value $ef$, regardless of the value of $V_g$. For a larger values of $A_{saw}$, such that $2\varepsilon_c>2eA_{saw}\geq\varepsilon_c$, the current will oscillate between the values $2ef$ and $3ef$, and so on. For negative values of $U_{sd}$, the current may be negative and oscillate between $-efn$ and $-ef(n+1)$ quantized values (where $n$ is an integer). $I_a(V_g)$ dependences similar to those in Fig.4c for different values of the source-drain bias (rather than for different values of $A_{saw}$ as in Fig.4c) were experimentally observed in a QD turnstile device [7, 8].

## IV. Experimental dependences of the acoustoelectric current.

We have shown that the ST model gives rise to very specific dependences on gate voltage and the source-drain bias. In this section we compare the predictions of the ST model with our experimental data, which mainly concerned the range of the RF power from 0 to 10 dBm, the interval where the best accuracy of the current quantization has been observed.

The authors of ref.1 refer to this interval as to high RF powers and state (we quote): "*For high rf powers, where the amplitude of the surface acoustic wave is much larger than the quantum dot energies, the quantized acoustoelectric current transport shows behaviour consistent with previously reported results. However, in this regime, the number of quantized current plateaus observed and the plateau widths are determined by the properties of the quantum dot, demonstrating that the microscopic detail of the potential landscape in the split-gate channel has a profound influence on the quantized acoustoelectric current transport*".

This statement forces us to re-examine the previous experiments [2-6, 11] in the context of the ST model. Another reason for doing that is that all the graphs in ref.1 that demonstrate the current quantization are identical in their appearance to the corresponding graphs obtained earlier [2-6, 11]. We believe that it is unlikely that the similar experimental graphs could be due to different mechanisms. The previous work contained more detailed (than in ref.1) investigation of the dependences of the current on source-drain bias and the gate voltage thus facilitating analysis of a possible relevance of the ST model.

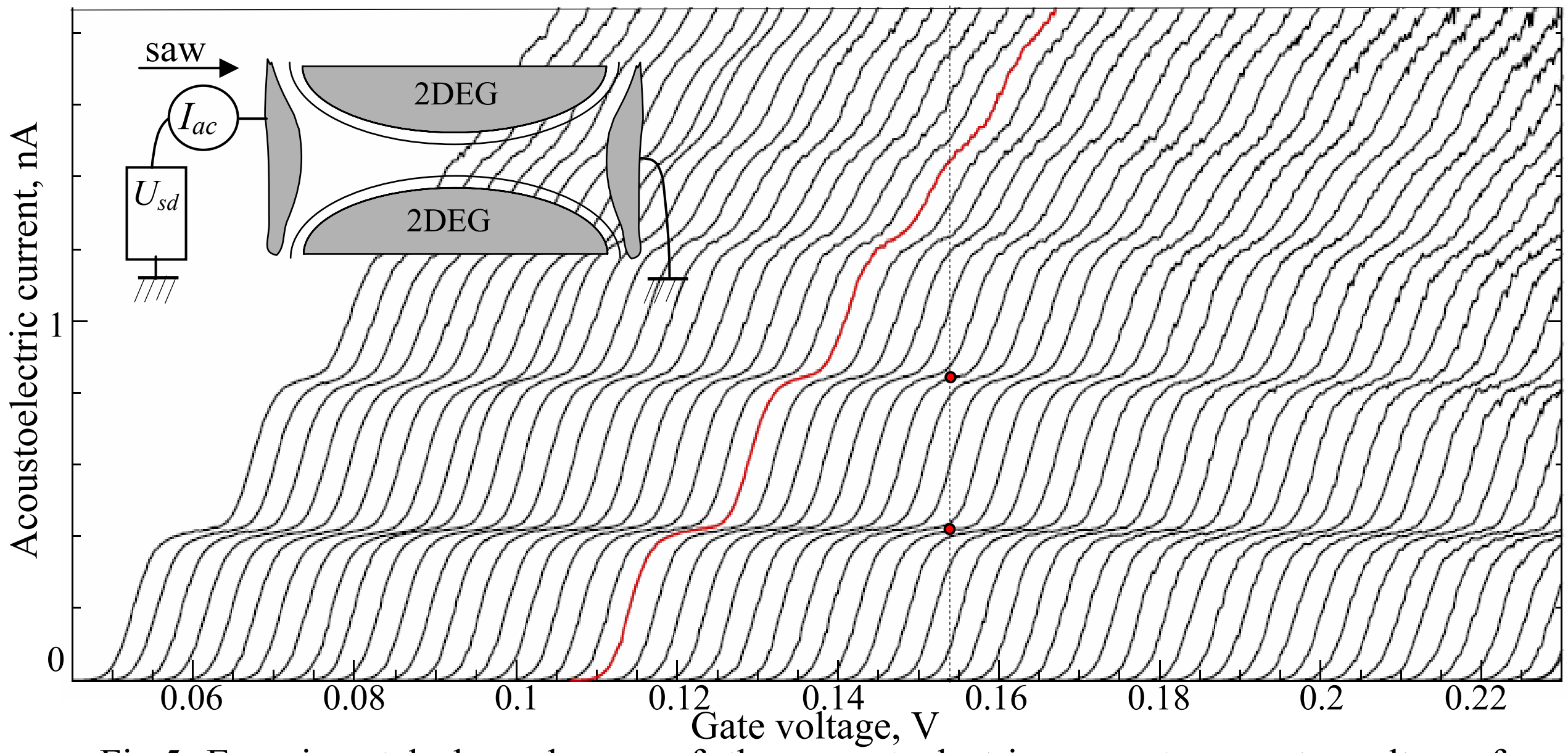

Fig.5 Experimental dependences of the acoustoelectric current on gate voltage for different values of source-drain voltage. Inset illustrates the design of the used one-dimensional channel. See text for details.

We shall demonstrate in this section that our experimental data [2-6, 11] are in profound disagreement with all the implications of the ST model, the disagreement which, in our opinion, eliminates the suggested in ref.1 link between the quantized plateaus and properties of a QD (in cases when there is one in the channel).

Representative experimental dependences (they were partly presented in ref. 11) of $I_{ac}(V_g)$ for different values of the bias are shown in Fig.5. The data were taken on a sample made from a GaAs-AlGaAs heterojunction with the electron density $2\times10^{11}$cm$^{-2}$ and the mobility ~$10^6$ cm$^2$V$^{-1}$s$^{-1}$. The quasi one-dimensional channel was defined by etched trenches so that the adjacent 2DEGs served as the gates (see ref.11 for details). The channel was depleted by the surface states in the trenches and the dc conductance pinch-off voltage was approximately +0.8V. The rf power applied to the SAW transducer was around 10 dBm, temperature 1.2K. No Coulomb blockade peaks were detected in the dc conductance for that sample.

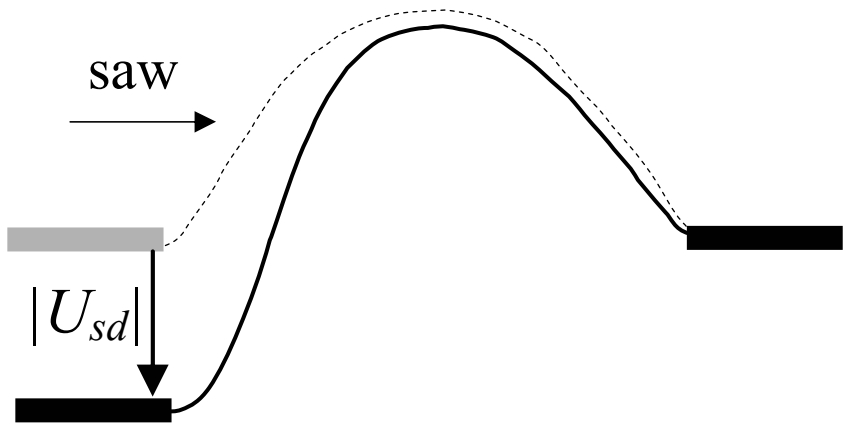

Fig.6

In Fig.5 the leftmost graph corresponds to —60 mV voltage applied to the source (entry) 2DEG with respect to the grounded drain 2DEG (see inset in Fig.5). In our notations it corresponds to the positive forward bias $U_{sd}$=+60 mV. The value of the voltage applied to the source 2GEG changes by +3mV for each consecutive graph up to the voltage of +111 mV (or the backward bias $U_{sd}$=—111 mV) for the rightmost graph. The red graph corresponds to zero bias. It is seen that the experimental $I_a(V_g)$ dependences in Fig.5 have nothing in common with oscillating $I_a(V_g)$ dependences inherent to the ST model (see Fig.4c). By considering values of the acoustoelectric current in Fig.5 at a fixed gate voltage and different values of the bias, it can be inferred that $I_a(U_{sd})$ dependences display a quantized current plateaux, with the centres of the subsequent plateaux values of $U_{sd}$ separated by approximately 12 mV in $U_{sd}$. This value follows from the observation that the segment of a vertical dashed line (in between the red dots) in Fig.5 that connects *ef* and *2ef* plateaux intersects four $I_a(V_g)$ graphs, resulting in 4×3 mV=12mV.

The $I_a(U_{sd})$ experimental dependence, that follows from Fig.5 is in disagreement with the ST model. We just point out the most apparent discrepancy with the ST model, namely the absence of the negative (that flows in opposite to the SAW propagation direction) acoustoelectric current under application of large backward bias. According to the ST model the acoustoelectric current can be estimated as $e^2f(2A_{saw}+U_{sd})/\varepsilon_c$. In Fig.5 the rightmost graph corresponds to $U_{sd}$~—110 mV, which is much larger than any plausible value of the $\varepsilon_c$ and larger than $2A_{saw}$ (within the ST model one has to accept $A_{saw}$~$\varepsilon_c$ in order to account for the presence of the first current plateau in the experimental graphs).

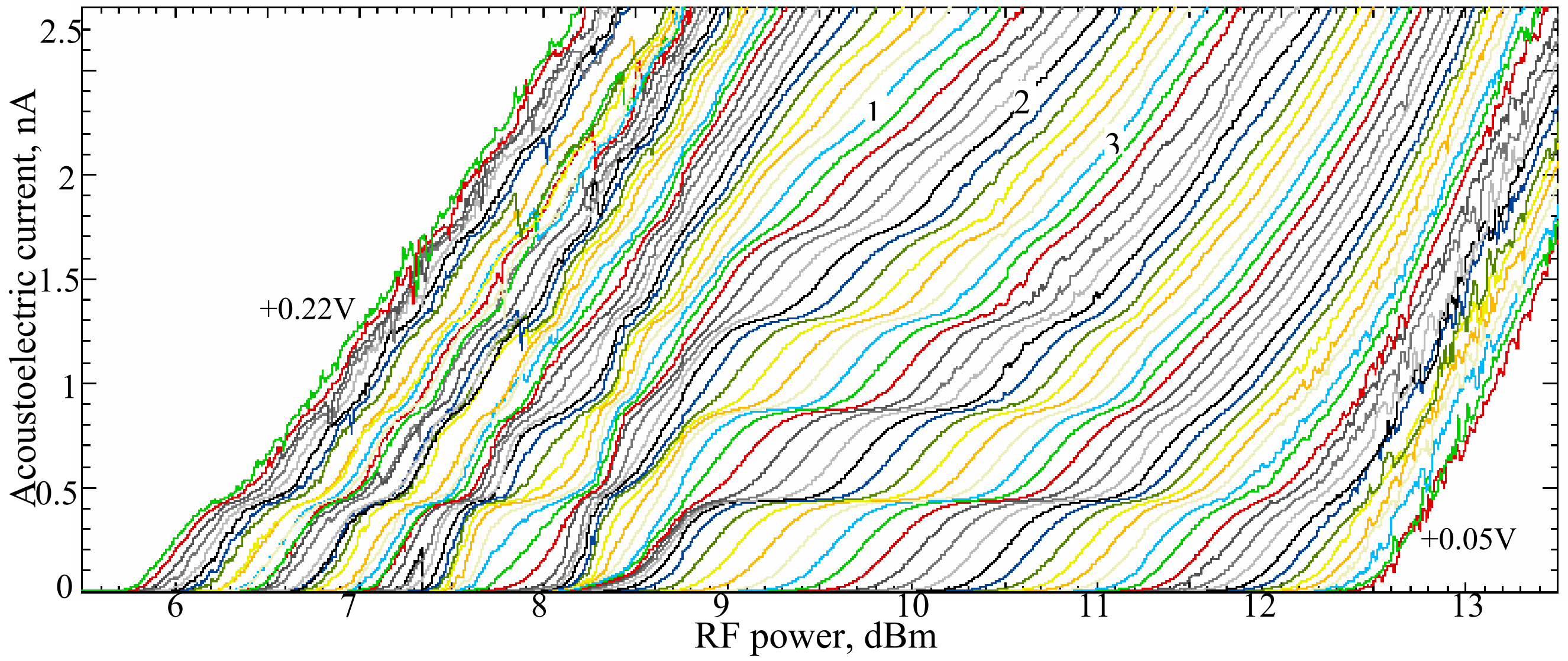

Fig.7 The acoustoelectric current as a function of the RF power for different values of gate voltage. See text for detail.

On the other hand the data in Fig.5 have very natural explanation in the terms of the original model [2, 3]. The source-drain bias changes the potential profile at the entrance to the channel. For instance, a backward bias makes the slope of the potential hill at the entrance steeper as shown in Fig.6. This change in the slope can be compensated by more positive gate voltage (that decreases the height of the potential hill) thus shifting the threshold for the acoustoelectric current (and a whole $I_{ac}(V_g)$ graph in Fig.5) towards more positive gate voltage values. Thus we conclude that the backward source-drain bias shifts $I_{ac}(V_g)$ graphs (for fixed SAW amplitude) to the right, and the forward bias shifts the graphs to the left, from the zero-bias graph. This is exactly the behaviour that Fig.5 demonstrates. Fig.7 presents (for the same sample) the dependences $I_{ac}(P)$ of the acoustoelectric current on the RF power for different values of the gate voltage in the interval from +0.05V (for the rightmost graph) up to 0.22V (for the leftmost graph), with the gate voltage increment of +2mV. It is seen that there are graphs with some plateaus missing. For instance graphs labelled by numbers 1, 2, and 3 have missing plateaux 3*ef*, 2*ef*, and *ef*, respectively. Most of the graphs have all four distinguishable plateaux, with the width of the subsequent plateaus decreasing. The ST model predicts quite different dependence on the power (see Fig.4a). On the basis of the ST model one would have expected the graphs $I_{ac}(P)$ displaying dominance of even or odd current plateaux, which is in apparent contradiction with the experimental data in Fig.7.

V. Has a new regime of the current quantization been really observed in ref.1?

The authors of ref.1 claim that (we quote): *"a regime of quantized transport is observed at low rf powers where the surface acoustic wave amplitude is comparable to the quantum dot charging energy. In this regime resonant transport through the single-electron dot state occurs which we interpret as turnstile-like operation in which the travelling wave amplitude modulates the entrance and exit barriers of the quantum dot in a cyclic fashion at GHz frequencies."*
The experimental evidence that authors of ref.1 gave in order to support their claim of the observation of a new regime of the quantized acoustoelectric current is presented in Figs. 3 and 6 (or Fig8a, which summarises the data in Fig.6) of ref.1. In Fig.3 of ref.1 the acoustoelectric current as a function of the rf power for different values of the gate voltage is shown. Fig.8a from ref.1 is reproduced here in Fig.8a (of the same number by coincidence) with the horizontal dashed lines and the question marks added by the author. Fig.8a is a grey-scale plot of the derivative $\partial I_{ac}/\partial A_{saw}$. The black lines schematically indicate area with large value of the derivative, and the light areas correspond to small values of the derivative. The dots in Fig.8a show number of electrons transferred across the channel per SAW cycle.

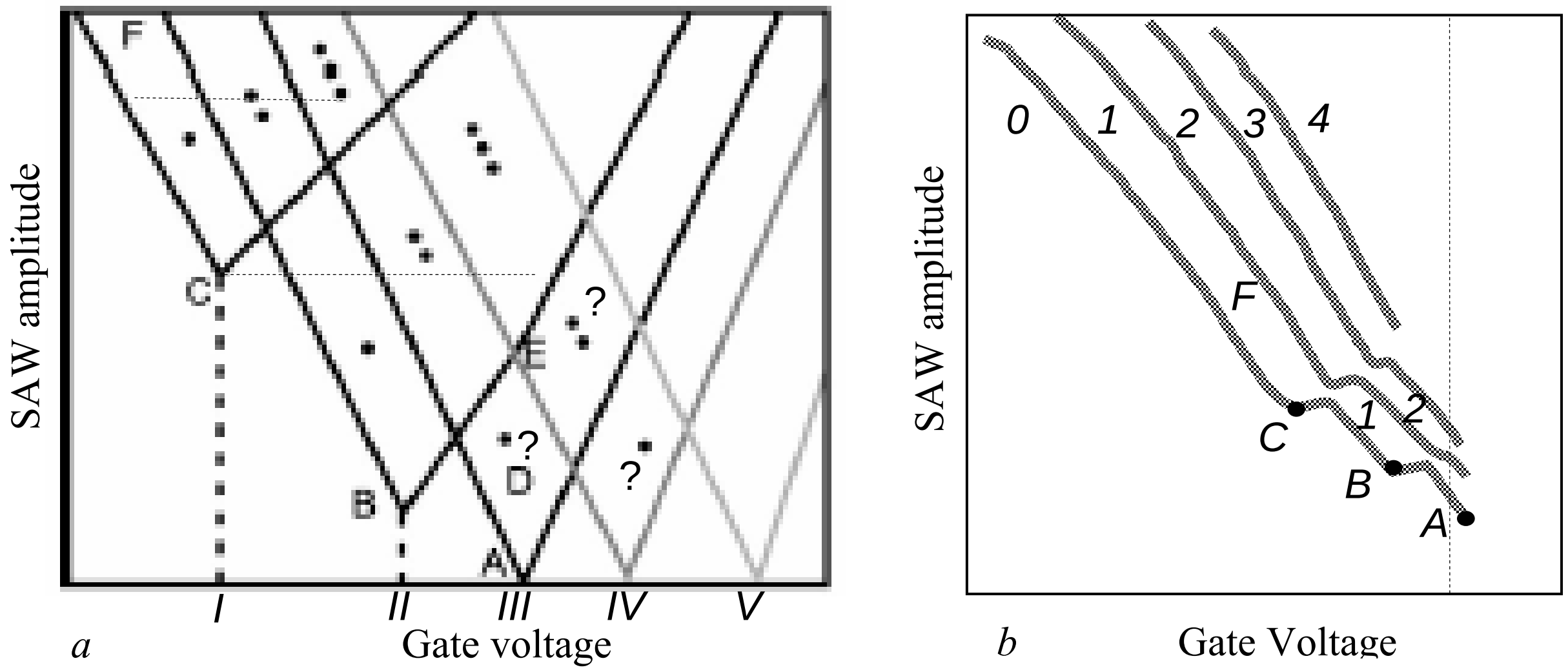

Fig.8 *a.* A grey-scale plot of the derivative $\partial I_{ac}/\partial A_{saw}$ as function of the SAW amplitude and gate voltage. The black areas indicate a larger value of the derivatives. *b.* the same as in panel *a* but in a wider area of the SAW amplitude and gate voltage. See text for detail.

The claim of ref. 1 is based on some features in the shape and positions of the domains in the ($A_{saw}$, $V_g$)-plane where the current quantization was (according to ref. 1) observed. Fig. 8a covers only a small part of the ($A_{saw}$ - $V_g$)-plane where the quantization of the current exists, and it is important (in the context of this Comment) to put the data in Fig. 8a into the prospective. This is done in Fig.8b where grey lines show the area with large derivative $\partial I_{ac}/\partial A_{saw}$, and the areas where quantized current is observed are marked by digits showing a number of electrons transferred per a cycle. The vertical dashed line indicates the pinch off voltage of the channel conductance. The area covered in Fig.8a constitutes a bottom right-hand part of Fig.8b, and the letters A, B, C and F correspond to the same values of $A_{saw}$ and $V_g$ in the both figures. In a standard situation when quantization of the current is observed without traces of an impurity induced QD in the channel, the grey-scale plot consists of a number of smooth lines like those in the top left-hand area of Fig.8b. When approaching the pinch off voltage the lines wash out because the quantization of the current ceases to exist in the open channel [2-6]. When an impurity is present nearby of the channel the acoustoelectric current graph acquires a characteristic peak structure in the gate voltage domain close to the conductance pinch off as had been established in refs.2-6. If both, the current quantization and the impurity-induced dot, are present, one could expect that the pattern of the smooth lines separating the quantization areas in the grey-scale plot will be distorted by the pattern due to the impurity. This view on the pattern in Fig.8b is in line with the suggestion [2, 4, 6], that an impurity potential hinders the current quantization.
The authors of ref.1 took a different point of view according to which the impurity defined QD in the channel causes the quantization of the acoustoelectric current. According to this point of view the smooth lines in the top left-hand part of Fig.8b are defined by the dot on the same basis as the impurity pattern close to the pinch off region.
They then claimed that at high rf powers: *"the number of quantized current plateaus observed and the plateau widths are determined by the properties of the quantum dot, demonstrating that the microscopic detail of the potential landscape in the split-gate channel has a profound influence on the quantized acoustoelectric current transport".*
We would like to stress that this statement (while presented in ref.1 as an established fact), in reality is nothing more than an assumption because it follows from the suggested (rather than proven) interpretation of the pattern of the grey-scale plot of the derivative $\partial I_{ac}/\partial A_{saw}$ in ref.1. We have demonstrated in the previous section that the ST model has no relevance to the current quantization observed in refs. [2-6], in the RF power interval from (approximately) 0 to 10 dBm.
The point of view accepted in ref.1 immediately raises at least 2 issues: (*i*) how to accommodate the situations when current reveals a very good quantization but no characteristic impurity peak structure is seen, and (*ii*) what is the mechanism of the current quantization by the QD. The former issue has not been addressed in ref.1, and for the resolution of the latter, the ST model was suggested. Now we would like to examine whether the experimental evidence in ref.1 support the claim concerning observation of a new regime of quantized transport through the single-electron dot state.

V.1 Dependence on the SAW amplitude.
ST model implies that a number of electrons transferred is determined by the ratio $2eA_{saw}/\varepsilon_c$. In Fig.3 of ref.1 the RF power range from -5 to 0 dBm corresponds to the range of electrostatic SAW potential peak-to peak amplitude ($2A_{saw}$ in our notations) from 0.8 to 1.5 mV. The Coulomb charging energy of the dot was determined in ref.1 to be equal $\varepsilon_c$=0.3 meV, so the interval 0.8 meV<$2eA_{saw}$<1.5 meV should contain at least 2 dot's energy levels. According to the ST model the *ef* acoustoelectric current plateau can't be observed in this power interval, which is in disagreement with the experimental data in Fig.3 of ref.1.
The horizontal dashed lines in Fig.8a correspond to the SAW amplitudes of 1.3 mV and 0.8 mV, respectively. According to the ST model, the current plateau *ef* is not allowed for the SAW amplitude of 0.8 mV (and the plateaux 1*ef* and 2*ef* are not allowed for the SAW amplitude of 1.2 mV), in apparent contradiction with the experiment.
We note that one can't escape the above conclusions on the basis of that values of $A_{saw}$ were not accurately estimated. The computation of the SAW amplitude in ref.1 was not independent of the ST model (the standard way is to calculate the SAW amplitude from the known geometry of the SAW transducer and the known value of the RF voltage applied to the transducer). Instead the authors of

ref.1 chose a procedure that prescribes the SAW amplitude a value which brings the ST model in agreement with experiment data in a small part of the pattern in Fig 6 of ref.1. It means that this procedure can't be changed when applied to other domains of the relevant parameters.

V.2 Dependence on the gate voltage.
By considering the current for a fixed value of the RF power it is easily seen from Fig.3 of ref.1 that the current vs gate voltage dependence is similar to that shown in Fig.6 of this Comment, rather than being of the oscillatory type as required by the ST model. The same conclusion follows from the data presented in Figs.6 and 8a of ref.1 and reproduced here in Fig.8a. Following along the horizontal dashed lines in Fig.8a, we see that the acoustoelectric current as a function of the gate voltage (at fixed SAW amplitude) reveals familiar (shown here in Fig.5) dependences which perfectly agree with the original model [2, 3] and in contradiction with the ST model.

V.3 Dependence on the source-drain bias.
Finally, we consider how the current in Figs. 3 and 6 of ref.1 depends on the source-drain bias. The authors if ref.1 wrote (we quote): *"In the regime where the quantum dot is well isolated from the leads (i.e., for moderate SAW amplitude and gate voltage), the current quantization and observed pattern of transitions is very robust. For applied source-drain bias voltages in the range of -1 mV* to*+1 mV (which is significantly larger than the quantum dot charging energy of 0.3 meV) no change in the pattern of current plateau transitions was observed".*
If we assume (unfortunately ref.1 does not specify range of the RF power and gate voltage to which the above statement applies) that the above statement applies to Fig.3 of ref.1 and to the part of Fig.6 of ref.1 that correspond to the pinched-off channel, then the dependence on the source-drain bias contradicts to the ST model as well. Changing the source-drain bias (at fixed SAW amplitude) from -1 mV to + 1 mV would have changed the acoustoelectric current by 6*ef* (see Fig.4b) for the $\varepsilon_c$=0.3 meV if the ST were relevant. We note that the source-drain bias provides a very simple and unambiguous means for the assessment of a turnstile model. If the dependences like those shown in Fig.4b of this Comment are not observed, the turnstile-type model should be discarded.
Thus the data presented in the main part of Fig.8a contradict to the ST model. The rest of the area covered by Fig.8a of ref.1 (and by Fig.8a of this Comment) concerns the open channel, or the domain of the gate voltages above the dc conductance pinch off. The crucial question here is whether the current quantization in the open channel was actually observed in ref.1. We have put the question marks in Fig.8a because our experiments have not demonstrated the quantization in the open channel. The quasi periodicity of the grey-scale pattern of the derivative $\partial I_{ac}/\partial A_{saw}$ in Figs. 6 and 8a of ref.1 does not constitute in itself the proof of the current quantization. Examination of our data (partly shown in Fig.2) concerning the impurity peak structure in the acoustoelectric current shows that they can lead to the grey-scale plot of the derivative $\partial I_{ac}/\partial A_{saw}$ similar to that in Fig.8a. However, it does not mean the current quantization in the units of *ef*, as Fig.2 illustrates.
Unfortunately the format (grey-scale plot) in which authors of ref.1 chose to present their data does not give the value of the current and therefore does not allow to check whether the quantized acoustoelectric current in the open channel had been indeed observed in ref.1. Neither this crucial piece of information could be extracted from the text of ref.1. The only remark that the authors give is (we quote): *"In the regime where the quantum dot is more strongly coupled to the leads (i.e., for low-SAW amplitude and gate voltage), normal source-drain bias driven conduction is observed which alters the absolute level of the SAW induced quantized current."*
It appears as that the demonstration of the quantization in the open channel (and the necessary subsequent proof of that the ST mechanism is responsible for that quantization) was substituted by the explanation of why the quantized current had not been observed.
In our view, the explanation is based on a wrong physical concept that the "*normal source-drain bias driven conduction*" could masks (by altering the absolute level of the SAW induced quantized current) "*the SAW induced quantized current*". A necessary condition for the turnstile mechanism to produce the quantized current is that in the absence of the RF signal, the current caused by a source-drain bias must be much smaller than *ef*. If this condition is fulfilled than the application of the source-drain bias will lead not to the "*normal source-drain bias driven conduction*" but will change the current in a very regular way as illustrated in fig.4, so that the quantization will be apparent. If however, the barriers are transparent when the SAW is off, then even in the absence of the bias, the operation of the ST device will not produce the quantized current because during some interval of the SAW cycle an electron can move from/in the dot in any direction rather than in the one prescribed by the ST model.

## Conclusions

The experimental graphs in ref.1 that unambiguously demonstrate the quantization of the acoustoelectric current are identical the graphs observed previously [2-6]. The graphs (including those in ref.1) are in contradiction with the turnstile-type (ST) mechanism [1] in terms of dependences on gate voltage, SAW amplitude, and source-drain bias.

In the domain of gate voltage close to the conductance pinch off, the current quantization had not been observed in the previous studies [2-6, 11]. Neither ref.1 presents the experimental evidence of the quantization in that domain.

We conclude that ref.1 has produced no experimental evidence of the quantized acoustoelectric transport through the quantum dot states.

## Acknowledgments.

This work was supported by the EC program "SAWPHOTON", the EPSRC grants GR/R54224, GR/R67521, GR/S59086, and the Newton Trust.

## References

1. N. E. Fletcher, J. Ebbecke, T. J. B. M. Janssen, F. J. Ahlers, M. Pepper, H. E. Beere, and D. A. Ritchie, Phys. Rev. B **68**, 245310 (2003).
2. J. M. Shilton, V. I. Talyanskii, M. Pepper, D. A. Ritchie, J. E. F. Frost, C. J. B. Ford, C. G. Smith, and G. A. C. Jones, J. Phys. Condens. Matter **8**, L531-L539 (1996).
3. V. I. Talyanskii, J. M. Shilton, M. Pepper, C. G. Smith, C. J. B. Ford, E. H. Linfield, D.A.Ritchie and G A C Jones, Phys. Rev. B **56**, 15180 (1997).
4. V. I. Talyanskii, J. M. Shilton, J. Cunningham, M. Pepper, C. J. B. Ford, C. G. Smith, E. H. Linfield, D. A. Ritchie, and G. A. C. Jones, Physica B, **249-251**, 140 (1998).
5. J. M. Shilton, PhD Thesis, Pembroke College, Cambridge, Chapter 6 (1997).
6. J. Cunningham, PhD Thesis, Selywn College, Cambridge, p.90 (2000).
7. L. P. Kouwenhoven, Phy. Rev. Lett. **67**, 1626 (1991).
8. L. P. Kouwenhoven, Thesis, Technical University of Delft, chapter 6, (1992).
9. C L Foden, V I Talyanskii, G J Milburn, M L Leadbeater, and M Pepper "High-frequency acousto-electric single-photon source" Phys. Rev. A **62**, 011803 (2000).
10. The use of the SAWs for quantum computation was proposed by J. Shilton in 1999. In the original Shilton's proposal a qubit was encoded by a position of an electron in two parallel 1D channels separated by a tunnel barrier. Later the Shilton's proposal was generalized to include the spin of an electron in a SAW dot in C. H.W. Barnes, J. M. Shilton, and A. M. Robinson, Phys. Rev. B **62**, 8410 (2000).
11. J. Cunningham, V. I. Talyanskii, J. M. Shilton, M. Pepper, A. Kristensen, and P.E. Lindelof, JLTP **118**, 555 (2000).